\documentclass[lettersize,journal]{IEEEtran}

\usepackage{amsmath,amssymb,amsfonts,amsthm}
\usepackage{xcolor}
\usepackage{multirow}
\usepackage[ruled,linesnumbered]{algorithm2e}
\usepackage{euscript}
\usepackage{makecell}
\usepackage{arydshln}
\usepackage{comment}
\usepackage{breqn}
\usepackage{bm}
\usepackage{caption}
\usepackage{hyperref}

\usepackage{soul}

\newtheorem{lemma}{Lemma}
\newtheorem{theorem}{Theorem}
\newtheorem{example}{Example}

\sethlcolor{white}

\begin{document}
\title{Efficient \hl{Additions and Montgomery Reductions} of Large Integers for SIMD}

\author{Pengchang Ren, Reiji Suda, and Vorapong Suppakitpaisarn~\IEEEmembership{Member,~IEEE} \thanks{P. Ren, R. Suda, and V. Suppakitpaisarn are with Department of Computer Science, Graduate School 
of Information Science and Technology, The University of Tokyo, Tokyo, Japan} 
\thanks{The work is partially supported by JSPS Grant-in-Aid for Transformative
Research Areas A grant number JP21H05845, and also by JST SICORP Grant Number JPMJSC2208, Japan. \hl{The authors would like to thank
anonymous reviewers and Taku Onodera for several useful comments and ideas which  significantly improved this paper.}}}

\maketitle

\begin{abstract}
This paper presents efficient algorithms, designed to leverage SIMD for performing Montgomery reductions and additions on integers larger than 512 bits. The existing algorithms encounter inefficiencies when parallelized using SIMD due to extensive dependencies in both operations, particularly noticeable in costly operations like ARM's SVE. To mitigate this problem, a novel addition algorithm is introduced that simulates the addition of large integers using a smaller addition, quickly producing the same set of carries. These carries are then utilized to perform parallel additions on large integers. For Montgomery reductions, serial multiplications are replaced with precomputations that can be effectively calculated using SIMD extensions. Experimental evidence demonstrates that these proposed algorithms substantially enhance the performance of state-of-the-art implementations of several post-quantum cryptography algorithms. Notably, they deliver a 30\% speed-up from the latest CTIDH implementation, an 11\% speed-up from the latest CSIDH implementation in AVX-512 processors, and a 7\% speed-up from Microsoft's standard PQCrypto-SIDH for SIKEp503 on A64FX.
\end{abstract}


\section{Introduction}
Prime field arithmetic forms the basis of many public-key cryptography algorithms, including RSA, ECDH, and various post-quantum cryptosystems such as SIKE \cite{NISTPQC-R4:SIKE22}, CSIDH \cite{AC:CLMPR18}, and CTIDH \cite{TCHES:BBCCLMSS21}.

The increasing availability of SIMD instructions on modern processors has highlighted the importance of utilizing these instructions to optimize the latency performance of large prime-field arithmetic. Recent research has shown that the latest support for 52-bit integer multiplication for AVX-512 on x64 can efficiently optimize prime field arithmetic for either throughput or latency \cite{VPMADD,TCHES:CFGRR21}. 

On the other hand, attempts to use ARM's Neon instructions have yielded little improvement for \hl{latency \mbox{\cite{NEONSIKE, anastasova2023time}}}.
Although Neon instructions have been used on ARM architecture, they have not resulted in significant latency improvement. This could be because Neon only allows for 32-bit integer multiplication and 128-bit SIMD, which may not be suitable for prime field arithmetic when dealing with primes as large as 512 bits. 

ARM has recently announced the introduction of a new SIMD extension known as SVE, designed to enhance their ARMv8 instruction set. SVE offers several improvements over Neon, including support for 64-bit integer multiplication and the ability to process larger vector lengths. In our prior research, we have enhanced the throughput of SIKE on SVE~\cite{SVESIKE}. Furthermore, to further optimize latency performance, Edamatsu et al. \cite{edamatsu2023efficient} have proposed a low-latency multiplication technique for large numbers specifically designed for SVE.

Previous research has primarily focused on optimizing the latency of multiplication operations using SIMD instructions and assumed that the algorithms for calculating addition and Montgomery reduction are already efficient. However, this assumption may not hold true for certain SIMD instructions, such as SVE, where additions and reductions can be expensive. This research has shown that these operations can become a bottleneck in various calculations, particularly in post-quantum cryptographic algorithms. Therefore, it is crucial to develop efficient algorithms that consider addition, multiplication, and reduction operations to fully leverage the benefits of SIMD instructions.

\subsection{Our Contributions}
The costs of additions and Montgomery reductions in SIMD are due to their long dependency chains. When using carry propagate adders, adding more significant limbs requires waiting for carries from less significant limbs before starting the addition. Similarly, Montgomery reductions are typically carried out limb by limb \hl{\mbox{\cite{montgomery1985modular,seo2016hybrid}}}. The results of the calculation of the less significant limbs are required for the calculation of the more significant ones. This research aims to break these dependency chains.

Section \ref{sec:add} introduces a novel addition algorithm that utilizes smaller additions to simulate the addition of large integers that require multiple limbs to store. The smaller additions give different results than the original large additions but produce the same set of carries. By using these carries, parallel additions on each limb of the large integers can be performed. For instance, the calculation of carries for an addition of 512-bit integers, requiring 8 limbs for storage in SVE, can be achieved by performing an addition of 64-bit integers, which requires only a single limb. The 64-bit addition can then serve as a basis to derive all the required carries for the original addition, enabling the parallel execution of additions across the eight limbs, predicated on the obtained carries.

Section \ref{sec:red} presents two techniques for Montgomery reduction. The first technique can be applied to any prime number, while the second is specifically designed for prime numbers of the form $2^\ell F - 1$. In the first technique, we notice that serial calculations in the reduction can be replaced with additions of independent multiplication results, which can be computed efficiently using SIMD. In the second technique,   it is demonstrated that reductions on such prime numbers can be performed using just two large multiplications. \hl{Then, by leveraging concepts from \mbox{\cite{TCHES:CFGRR21, edamatsu2020accelerating}}, the Karatsuba algorithm is utilized to speed up these large multiplication operations.}

Lastly, experimental results for the proposed techniques are shown in Section \ref{sec:results}.
For addition, benchmark results show that the proposed algorithm is faster than the carry propagation addition implemented in AVX-512 by up to $2.5$ times. The proposed SIMD addition is faster than the carry propagation addition implemented in x64 by up to $32\%$.

The proposed techniques for Montgomery reduction have shown significant speed improvements over previous implementations. Benchmark results demonstrate a 11\% speedup for Montgomery multiplication and a 36\% speedup for Montgomery squaring, compared to the state-of-art implementation in AVX-512 implementation by Cheng et al. \cite{TCHES:CFGRR21}. Compared to x64 implementation, the improvement is 97\% for Montgomery multiplication and 151\% for Montgomery reduction. 

Furthermore, the proposed techniques can accelerate the calculation of CSIDH by 11\% compared to the work by Cheng et al.~\cite{TCHES:CFGRR21}, and an 77\% compared to x64 implementation. These results demonstrate the effectiveness of the proposed techniques for improving the efficiency of post-quantum cryptographic algorithms.

In addition to the implementation by Cheng et al.~\cite{TCHES:CFGRR21}, it can be demonstrated that the proposed additions and Montgomery reductions can significantly reduce the computation of several post-quantum cryptography algorithms. Particularly, a 30\% increase in speed can be achieved compared to the latest implementation of CTIDH by Benegas et al. \cite{TCHES:BBCCLMSS21} in SVE. Moreover, the proposed Montgomery reduction for the prime number $p_{503} = 2^{250}3^{159} - 1$, which is the standard prime used for SIKE, proves to be 26\% faster than the reduction in Microsoft's standard PQCrypto-SIDH. This contributes to an overall 7\% improvement in computation for SIKE.

\section{Preliminaries}
\label{sec:preliminaries}
\subsection{Large Integer Representation}
The implementation of prime field arithmetic requires careful consideration of how to store large integers in memory. An approach is to use radix-$2^\omega$, where a large integer is stored using several $\omega$-bit integers. One $\omega$ bit integer is called a \textit{limb}. If $\omega$ is equal to the machine word size, this is referred to as using the \textit{native radix}, while if $\omega$ is smaller, it is called \textit{reduced radix}. Throughout this paper, the symbol $n$ is used to represent the number of limbs required to store a large integer.

Several algorithms \hl{such as \mbox{\cite{VPMADD,TCHES:CFGRR21}}} employ register packing of multiple large integers to increase throughput. Suppose a large integer requires 8 limbs to store and a single register can hold up to 8 limbs. It is possible to use one register to store all 8 limbs of the integer. Alternatively, it is possible to utilize two registers to pack two integers, storing four limbs of each integer in each register. The representation of the packing of $x$ integers together is referred to as $x$-\textit{way packing}. 

The optimal values of $x$ and $\omega$ can vary depending on the architecture and prime field being used, and selecting them is outside the scope of this investigation.

\subsection{Single Instruction Multiple Data (SIMD)}
\textit{SIMD instructions} are a class of instructions that enable the manipulation of multiple data elements in a single instruction. They are found in modern CPUs, including AVX-512 on x64, Neon, and SVE on ARM. SIMD instructions offer different levels of support for various calculations and \hl{different} vector length. For instance, AVX-512 has a vector length of 512 bits, Neon's vector length is 128 bits, while SVE ranges from 128 to 2048 bits. 

A single vector has the capacity to store multiple integers, with each integer occupying a specific section of memory referred to as a \textit{lane}.  AVX-512 register can be interpreted as 64 8-bit lanes, 32 16-bit lanes, 16 32-bit lanes or eight 64-bit lanes. With most SIMD instructions, performing an operation using SIMD means applying that operation to all lanes of vectors simultaneously. For example, adding two vectors involves adding each lane of the two vectors independently. \hl{In this study, the terms SIMD\_ADD and  SIMD\_SUB denote the process of carrying out parallel addition and subtraction operations using SIMD technology. For multiplication operations, when two one-limb operands yield a two-limb result, the terms SIMD\_MUL\_H and SIMD\_MUL\_L are employed. These denote the process of calculating the more and less significant limbs of the outcome, respectively.}
In contrast, there are instructions designed to move data across lanes, like the table lookup instruction TBL in SVE. These are referred to as cross-lane instructions and are generally more computationally expensive. 

Intel CPUs from the 10th generation and later have extensively incorporated AVX-512. While Neon is available on numerous modern ARM CPUs, SVE with 512-bit vectors is now exclusively available on CPUs intended for high-performance computing, such as A64FX.

\begin{table*}[t]
\caption{Instruction latency \hl{in clock cycles} and Cycles per instruction (CPI) for x64 instructions on Intel Tigerlake, AVX-512 instructions on Intel Tigerlake, \hl{A64 instructions on Fujitsu A64FX}, and SVE instructions on Fujitsu A64FX}
\label{tab:lat}
\begin{center}
\begin{tabular}{c||cc|cc||c|c}
\hline
 \multirow{2}*{\textbf{Operation}}&\multicolumn{2}{c|}{\textbf{Tigerlake x64}} &\multicolumn{2}{c||}{\textbf{Tigerlake AVX-512}} &\textbf{A64FX A64}&\textbf{A64FX SVE}\\ 
 ~&Latency&CPI&Latency&CPI&Latency&Latency\\
 \hline
 Cache Access&3&0.5&4&0.5&5&11\\
Addition&1&0.25&1&0.5&1&4\\
Logic&1&0.25&1&0.5&1&4\\
Shift&1&0.5&1&1&2&4\\
Compare&1&0.25&3&1&1&4\\
Popcount&3&1&3&1&-&4\\
Multiplication, 64-bit&3&1&-&-&5&9\\
Multiplication, 52-bit&-&-&4&1&-&-\\
Table lookup/Cross-lane&-&-&3&1&-&6\\
\hline
\end{tabular}
\label{tab:inst}
\end{center}
\end{table*}

Table \ref{tab:inst} presents latency in cycle times and throughput metrics as cycles per instruction (CPI) for both Tigerlake's x64 and AVX-512. It also includes latency figures for A64 and SVE on Fujitsu's A64FX \cite{fog2022instruction,Fujitsu}. \hl{Latency denotes the time required to complete a single computation for the instructions, while throughput is defined as latency over the number of instructions that can be processed simultaneously. Throughput data is typically considered when executing several instructions concurrently, whereas latency data applies to tasks that cannot be carried out in parallel.} Regrettably, Fujitsu has not supplied throughput information for each instruction, resulting in the listing of only latency data.

Avoiding instructions with high latency is desirable for any program. Table \ref{tab:inst} demonstrates that cache access incurs a significant cost for all instruction sets listed, highlighting the need to minimize data movement between SIMD registers and general purpose registers.  This is also true for cross-lane instructions.

The table also highlights that all SVE instructions on A64FX exhibit high latency. 
Typically, when optimizing calculations, the number of additions and comparisons is not a major concern as they are not as expensive as other operations. However, in the case of SVE, their cost is significant and cannot be ignored.

\subsection{Isogeny-based Cryptography: CSIDH, CTIDH, and SIKE}

CSIDH \cite{AC:CLMPR18} is an isogeny-based cryptography algorithm developed by Castryck et al. While the original implementation is not constant-time, which could lead to timing attacks, several attempts have been made to find a fast constant-time implementation \cite{ACNS:HLKA20,COSADE:JAKJ19,EPRINT:OAYT19,LC:CCCDRS19}. Unfortunately, according to the discussion in Cheng et al. \cite{TCHES:CFGRR21}, these implementations are significantly slower than the implementation which is not constant-time. CTIDH is a new algorithm developed by Banegas et al. \cite{TCHES:BBCCLMSS21} that improves constant-time performance, with over 60\% speed-ups achieved by changing the key space of CSIDH.

Both CSIDH and CTIDH are defined over a finite field $\mathbb{F}_p$, where $p=(4\cdot \prod l_i )- 1$ and $l_i$-s are small odd primes. Although both algorithms offer various primes to meet different security levels, most optimizations for CSIDH have been discussed for primes with 511 bits \cite{TCHES:CFGRR21, ACNS:HLKA20,COSADE:JAKJ19,EPRINT:OAYT19,LC:CCCDRS19}. Most source codes for CSIDH and CTIDH are written in x64 assembly. Up to our knowledge, only implementation by Jalali et al.~\cite{COSADE:JAKJ19} is on ARM architecture. A constant-time implementation of CSIDH for ARM architecture was proposed in the same work, but it was computationally intensive. Currently, there is no implementation for CTIDH on ARM architecture.

SIKE \cite{NISTPQC-R4:SIKE22} is another isogeny-based cryptosystem based on Supersingular Isogeny Diffie-Hellman (SIDH) key exchange \cite{PQCRYPTO:JaoDeFo11}. It was one of the most well-known isogeny-based cryptosytem, and has been proposed as one of the four alternate candidates in the fourth round of NIST's PQC standardization process. However, in August 2022, it was reported that an algorithm \cite{EPRINT:CasDec22}  could attack the current implementation of SIKE, and subsequent research showed that minor changes do not make the protocol secure \cite{EPRINT:MaiMar22,EPRINT:Robert22}. 
\hl{While none of the algorithms  proposed in this paper are exclusive to SIKE, but, as the implementations of SIKE were highly optimized, they are used as benchmarks of this work.}

The protocol of SIKE is defined over the quadratic extension of the finite field $\mathbb{F}_{p^2}$, where $p=2^{e_1}3^{e_2} -1$ and $2^{e_1}\approx 3^{e_2}$. It defines four parameter sets, with prime bit lengths of 434, 503, 610, and 751. Microsoft has an implementation of SIKE called PQCrypto-SIDH \cite{PQCryptoSIDH}, optimized for x64 and ARM architecture using handwritten assembly. The reference implementation used in this paper is SIDHv3.5.

\section{Previous Approachs on Additions and Montgomery Reductions}

\subsection{Carry-Propagate Addition}
\label{sec:carryProp}

One way to implement addition is through digit-by-digit carry propagation. The addition is calculated for each limb, starting from the least significant limb to the most significant limb, while considering the carry. Subtraction can be carried out using a similar approach. 

General purpose instructions can calculate the addition between two large integer $A$ and $B$ efficiently. For $A,B$ stored in radix $2^\omega$, denote the $i$-th least significant limb of $A$ and $B$ by $A_i$ and $B_i$, i.e., ${A=\sum^{n-1}_{i=0}2^{i \cdot \omega}A_i}$,  and ${B=\sum^{n-1}_{i=0}2^{i \cdot \omega}B_i}$. The carry propagation addition of $A$ and $B$ can be simply implemented with repeatedly calculating $ {D_i\gets A_i+B_i+c_i \bmod 2^\omega}$ and ${c_{i+1}\gets\lfloor (A_i+B_i+c_i)/2^\omega\rfloor}$, starting with $c_0=0$. 
 
Unfortunately, the carry propagation algorithm performs poorly in SIMD. In order to obtain a value $c_i$ for the $(i + 1)$-th limb addition, all less significant limbs must be added first, creating a long dependency chain that cannot be parallelized. Additionally, the result $c_{i+1}$ is stored in the $i$-th lane of the SIMD register, but it is needed in the $(i+1)$-th lane, requiring an expensive cross-lane operation unless each SIMD register only contains one limb. These issues are severe when implementing on A64FX, where all SVE instructions have high latency, as shown in Table~\ref{tab:inst}. 

\subsection{Carry-Select Addition~\mbox{\cite{bedrij1962carry}}}
\label{sec:carrySec}

The carry-select addition approach was introduced to circumvent the long dependency chain in carry-propagate addition. In carry-propagate addition, it is necessary to wait for the value of $c_i$ before computing the sum $D_i = A_i + B_i + c_i$. Rather than waiting for the availability of $c_i \in \{0,1\}$, carry-select addition computes the sum results for both potential values of $c_i$ in advance. That is, the adders calculate $D_i^{(0)} = (A_i + B_i) \bmod 2^\omega$, $D_i^{(1)} = (A_i + B_i + 1) \bmod 2^\omega$, $c_{i + 1}^{(0)} = \lfloor (A_i + B_i)/2^\omega \rfloor$, and $c_{i + 1}^{(1)} = \lfloor (A_i + B_i + 1)/2^\omega \rfloor$ for $i > 1$. Since $D_0 = (A_0 + B_0) \bmod 2^\omega$ and $c_1 = \lfloor (A_0 + B_0)/2^\omega \rfloor$, it becomes feasible to use the $c_i$ value from the less significant limb to select the value of $D_i$ and $c_{i + 1}$ at the more significant limb, meaning $D_i = D_i^{(c_i)}$ and $c_{i + 1} = c_{i + 1}^{(c_i)}$.

With this approach, it becomes feasible to eliminate additions from the lengthy dependency chain through the pre-computation of $D_i^{(0)}$, $D_i^{(1)}$, $c_{i + 1}^{(0)}$, and $c_{i + 1}^{(1)}$. However, this elimination needs the computation of results for both $c_{i}$ values, effectively doubling the calculation effort. Furthermore, the long dependency chain continues to persist within the selection process.  

\subsection{Kogge-Stone Vector Addition}
\label{sec:Kogge}

\hl{An approach presented in \mbox{\cite{avxadd}}, based on the Kogge-Stone vector addition technique \mbox{\cite{kogge1973parallel}}, demonstrates high efficiency in the AVX-512 architecture.} This approach uses specific AVX-512 instructions to compute all $c_i$ values concurrently, thereby mitigating the lengthy dependency chain found in the carry-select adder.

We will not discuss the specifics of this approach due to its reliance on unique AVX-512 instructions, which limits its applicability to other architectures. Moreover, these instructions yield $c_i$ values on a register called mask registers \footnote{While SVE refers to mask registers as predicate registers, we have opted to maintain consistency in terminology within this paper. Consequently, predicate registers in SVE will also be referred to as mask registers to ensure straightforward understanding.}. This requires a transfer of these values to general-purpose registers. Although this transfer operation is relatively quick on AVX-512, it proves significantly more time-consuming on SVE.

Nevertheless, we use this approach as a benchmark for the AVX-512 architecture. We also attempt to implement this approach on SVE, using it as our benchmark in this respective architecture.

\subsection{Montgomery Reduction \cite{montgomery1985modular}}
\label{sec:mont}

Let $R = 2^{\omega n}$, and let $R^{-1}$ be an integer that satisfies $R\cdot R^{-1} \equiv 1 \pmod{p}$. Define a \textit{Montgomery reduction} of $T$ when $T < pR$ as $\textit{REDC}(T)$, which calculates $TR^{-1} \bmod p$. Additionally, define a partial Montgomery reduction function, denoted as \textit{redc}$(T,k)$ for $0 < k \le \omega n$, as a function which returns a positive integer such that $\textit{redc}(T,k) \equiv T2^{-k} \pmod{p}$. While calculating modular reduction, multiplying $T$ with $R^{-1}$ and performing the modulo operation is a straightforward method. However, the modulo operation can be computationally expensive. Avoidance of this situation can be achieved by employing the technique detailed in Algorithm~\textsf{ExistingGenericRedc}. \hl{In the algorithm, the symbol $T^{(i)}$ is used to represent the partial reduction $redc(T,\omega i)$.}

\begin{algorithm}
 \SetAlgorithmName{Algorithm~\textsf{ExistingGenericRedc}}{}{}
 \caption{Generic Montgomery reduction}
 \KwIn {$T < pR$, where $p  < 2^{\omega n}, r=2^\omega, R=2^{\omega n} $, $r^{-1}, R^{-1}$ are positive integers such that $rr^{-1} \equiv 1 \mod p$, $RR^{-1} \equiv 1 \mod p$, \hl{and $p'\gets \frac{rr^{-1}-1}{p}$}}
 \KwOut  {$REDC(T) = TR^{-1} \bmod p$}
  $T^{(0)} \gets T$ \\
 \For {$i = 1$ \KwTo $n$}{
$Q\gets T^{(i - 1)}p'\bmod r$ \\
 $T^{(i)}\gets (T^{(i-1)} +Qp)/r$\\
  }
  \uIf{$T^{(n)} >p$}{$T^{(n)} \gets T^{(n)}-p$}
 \Return $T^{(n)}$ 
 \end{algorithm}

To explain Algorithm~\textsf{ExistingGenericRedc}, notice that the loop in the algorithm computes the partial reduction of $T^{(i - 1)}$, i.e., $T^{(i)} \equiv T^{(i-1)}2^{-\omega} \mod p$. The correctness of this partial reduction can be proven as follows. By having $r = 2^\omega$, $m = (rr^{-1} - 1)/p$, and $q = T^{(i - 1)}m \mod r$ (Line 3):
\begin{dmath*}
T^{(i - 1)} + Qp \equiv T^{(i - 1)} + T^{(i - 1)}p'p \equiv T^{(i - 1)}(1 + rr^{-1} - 1)
\equiv T^{(i - 1)}rr^{-1} \equiv 0 \pmod{r}
\end{dmath*}
Hence, $T^{(i - 1)} + Qp$ is divisible by $r$. Because of that, $T^{(i)}$ are integers for all $i$. Also, because $T^{(i-1)} \equiv T^{(i - 1)} + Qp \pmod{p}$, we obtain $T^{(i)} \equiv T^{(i - 1)} r^{-1} \pmod{p}$ as expected. 

As a result from the previous paragraph, 
$T^{(i)} \hiderel{\equiv} T^{(0)} r^{-i} \equiv T r^{-i} \pmod{p}.$
Hence, 
$T^{(n)} \hiderel{\equiv} T r^{-n} \equiv TR^{-1} \pmod{p}.$
To show that $T^{(n)} = \textit{REDC}(T) = TR^{-1} \bmod p$, it is only left to show that $T^{(n)} < p$. Indeed, it can be shown by induction that $T^{(i)} < pr^{n - i} + p$, because $T^{(0)} < pR = pr^{-n}$ and
\begin{dmath}
    T^{(i)} = \frac{T^{(i - 1)} + Qp}{r}  <  \frac{pr^{n - i + 1} + p + (r - 1)p}{r} = \frac{pr^{n - i + 1} + rp}{r} \hiderel{=} pr^{n-i} + p. \label{eqn:mont0}
\end{dmath}
Hence, $T^{(n)}$ obtained from the loop is smaller than $pr^{n - n} + p = 2p$. By Lines 6-8, the returned value is smaller than $p$.

By utilizing the \textit{REDC}$(T)$ function calculated in Algorithm~\textsf{ExistingGenericRedc}, computing the modulo $p$ can be avoided. 
Instead, the reduction can be calculated by a series of modulo and division operations using $r = 2^\omega$. Since all the integers are stored in binary, modulo and division by $r$ can be computed by discarding some of the binary words if radix-$2^\omega$ is used. 

\subsection{Montgomery Representation and Multiplication}

In this paper, elements are taken from the prime field $\mathbb{F}_p = \{0, \dots, p-1\}$. The multiplication of $A,B \in \mathbb{F}_p$ is computed as $A \cdot B \bmod p$, where the multiplication can be done efficiently, but the modulo operation afterwards can be time-consuming. Montgomery reduction in the previous subsection helps reducing the computation of the multiplication~\cite{montgomery1985modular}.

Recall from the previous subsection that $R = 2^{\omega n} > p$. Since $p$ is an odd number, it is coprime to $R$. Therefore, for any $A, B \in \mathbb{F}_p$ with $A \neq B$, $AR \not\equiv BR \pmod p$. Hence, $A$ can be represented with another representation $\widehat{A} = AR \mod p$, known as the \textit{Montgomery representation} of $A$.

The application of the Montgomery reduction to $\widehat{A}\widehat{B}$ enables the calculation of the Montgomery representation of $AB$. This is because $\textit{REDC}(\widehat{A}\widehat{B}) = (AR)(BR)R^{-1} \bmod p = (AB)R \bmod p$. This approach allows us to perform the reduction operation following each multiplication rather than incurring the costly modulo operation after every multiplication. When numbers are represented using Montgomery representation, this approach can substantially reduce the cost of multiplication in prime field arithmetic since the reduction operation is much less expensive than the modulo operation. The multiplication of two numbers in the Montgomery representation is referred to as \textit{Montgomery multiplication}.

\subsection{More Efficient Montgomery Reduction for Primes in the Form $2^\ell F - 1$ \cite{faz2017faster}}

A prime number $p$ is considered to be $\lambda$-\textit{Montgomery-friendly} if it satisfies the condition $p\equiv \pm 1\bmod 2^{\lambda \cdot \omega}$, where $\lambda$ and $\omega$ are positive integers. Montgomery reduction can be accelerated when a prime number $p$ satisfies these specific conditions. An efficient algorithm for performing Montgomery reduction on $\lambda$-Montgomery-friendly moduli is proposed by Faz-Hern\'{a}ndez et al. and presented as Algorithm~\textsf{ExistingSpecificRedc} \cite{faz2017faster}. It has been demonstrated to be significantly faster than Algorithm~\textsf{ExistingGenericRedc}.

 \begin{algorithm}
 \SetAlgorithmName{Algorithm~\textsf{ExistingSpecificRedc}}{}{}
 \caption{Montgomery reduction with $\lambda$-Montgomery-friendly modulus \cite{faz2017faster}}
 \label{alg:mont1}
 \KwIn {$p  < 2^{\omega n}, r=2^\omega , R=2^{\omega  n}, T < pR, 1 < m \leq \lambda$
 such that $p \bmod r^m \equiv -1$, $\lambda_0\gets \lfloor \omega n/ m \rfloor$, $\lambda'_0\gets  (\omega \cdot n)\bmod m$, and $M\gets  (p+1)/2^{\lambda \cdot \omega}$.}
 \KwOut  {$TR^{-1} \bmod p$}
 $T^{(0)} \gets T$\\
 \For{$i\gets 1$ \KwTo $\lambda_0$}{
 $Q\gets T^{(i - 1)} \bmod 2^{m\cdot \omega}$\\
 $T^{(i)}\gets \left\lfloor (T^{(i - 1)}+2^{\lambda\cdot \omega}Q\cdot M)/2^{m\cdot \omega}\right\rfloor$
 }
 \If{$\lambda'_0\ne 0$}{
  $Q\gets T^{(\lambda_0)} \bmod 2^{\lambda'_0\cdot \omega}$\\
 $T^{(\lambda_0 + 1)} \gets \left\lfloor (T^{(\lambda_0)}+2^{\lambda\cdot \omega}Q\cdot M)/2^{\lambda'_0\cdot \omega}\right\rfloor$
 }{}
  \If{$T^{(\lambda_0 + 1)}\ge p$}{$T^{(\lambda_0 + 1)}\gets T^{(\lambda_0 + 1)}-p$}{}
 \Return $T^{(\lambda_0 + 1)}$ 
 \end{algorithm}
 
Many cryptographic systems, such as SQISign \cite{EPRINT:DeFLerWes22}, CGL hash function \cite{EPRINT:DolPerBar17}, and SIKE \cite{NISTPQC-R4:SIKE22}, use prime numbers of the form $p = 2^{\ell}F - 1$ as moduli. For any $\lambda \leq \ell / \omega$, these numbers are in the form $p \equiv -1 \bmod 2^{\lambda \cdot \omega}$, and therefore, are $\lambda$-Montgomery-friendly. For the simplicity of our explanation, we assume that $p \equiv -1 \mod 2^{\lambda \cdot \omega}$ or $p + 1$ is divisible by $2^{\lambda \cdot \omega}$ in the remaining part of this subsection.
The value of $M$ is then a positive integer. 


We can express $r^{-1}$ for a given $p$ as $r^{-1} = 2^{\ell - \omega}F$, since $$2^\omega 2^{\ell - \omega}F \equiv 2^\ell F \equiv 1 \pmod{p}$$ for such $p$.
The value of $p' = \frac{rr^{-1} - 1}{p}$ in Algorithm~\textsf{ExistingGenericRedc} can be computed as: $$\frac{rr^{-1} - 1}{p} = \frac{2^\ell \cdot F - 1}{p} = \frac{p}{p} = 1.$$ \hl{This means that the multiplication by $p'$ on Line 3 of Algorithm~\textsf{ExistingGenericRedc} can be avoided, as shown in Line 3 of Algorithm~\textsf{ExistingSpecificRedc}.} 

Another improvement of this algorithm arises from the fact that $Q$ is multiplied by $p+1=2^{\ell}F$ instead of $p$ at Line 4 of the algorithm. Since the least significant bits of $p+1=2^{\ell}F$ are zeros, the integers involved in this multiplication are smaller than those at Line 4 of Algorithm~\textsf{ExistingGenericRedc}, resulting in a faster multiplication. Recall that $T^{(i - 1)} + Qp$ is divisible by $r = 2^{m\omega}$. The result remains the same even after replacing $p$ with $p+1$, because
\begin{dmath*}
    \left\lfloor \frac{T^{(i-1)} + Q(p + 1)}{r} \right\rfloor = \frac{T^{(i-1)} + Qp}{r}+ \left \lfloor\frac{Q}{r} \right\rfloor = \frac{T^{(i-1)} + Qp}{r}
\end{dmath*}

There is also a technique called lazy correction for Montgomery reduction. The final correction in Algorithms \textsf{ExistingGenericRedc} and \textsf{ExistingSpecificRedc} can be avoided if $R>4p$. It is because of the fact that if $\widehat{A}, \widehat{B} <2p$ and $R>4p$, then $T^{(n)}$ at Line 6 of Algorithm~\textsf{ExistingGenericRedc} and $T^{(\lambda_0 + 1)}$ at Line 10 of Algorithm~\textsf{ExistingSpecificRedc} is less than $2p$ \cite{EPRINT:Trei13}. By performing all field arithmetic in $\mathbb{F}_{2p}$ instead of $\mathbb{F}_{p}$, the final correction is only needed in rare cases, such as when performing an equality test or generating the result.

\section{Proposed SIMD Addition}
\label{sec:add}

\hl{We propose an addition and carry propagation for SIMD in Algorithm~\textsf{ProposedAdd}.  The algorithm uses some ideas from the carry-select adder~\mbox{\cite{bedrij1962carry}}, which is described in Section {\ref{sec:carrySec}}. To eliminate the long dependency chain associated with carry selection, the algorithm simulates the addition of larger integers via the summation of smaller ones that fit within a single limb. As a result, all the carries are obtained in the general-purpose registers, thus bypassing the need for transferring data from mask registers to general-purpose ones as required by the approach discussed in Section {\ref{sec:Kogge}}.}

\begin{algorithm}
\SetAlgorithmName{Algorithm~\textsf{ProposedAdd}}{}{}
\captionsetup{labelformat=empty} 
 \caption{Proposed SIMD addition}
 \label{alg:carryprop}
 \KwIn {integer $A, B$ in radix-$2^\omega$, where $A=\sum^{n-1}_{i=0}2^{i \cdot \omega}A_i$, $B=\sum^{n-1}_{i=0}2^{i \cdot \omega}B_i$, and $A_i, B_i <2^\omega$} 
 \KwOut  {$A+B$}
 \tcc{$\mathsf{s}_i,\mathsf{t}_i$ and $\mathsf{p}_i$ are 8-bit integer.
$A_i, B_i, D_i$ and $G_i$ are \hl{$\omega$}-bit integer.
$c_i$, $m_i$ are 1-bit integer.
}
 $\langle D_i \rangle_{i = 0}^{n - 1} \gets \text{SIMD\_ADD}\left(\langle A_i \rangle_{i = 0}^{n - 1}, \langle B_i \rangle_{i = 0}^{n - 1}\right)$ \\
 Use SIMD to calculate $\mathsf{t}_i$ and $\mathsf{p}_i$ as in Table \ref{tab:cond} in parallel. \\
   $(\mathsf{s}_{n-1}, \dots, \mathsf{s}_0) \gets (\mathsf{t}_{n-1}, \dots, \mathsf{t}_1) + (\mathsf{p}_{n-1}, \dots, \mathsf{p}_1)$ \\ 
 $\langle \mathsf{s}_i \rangle_{i = 0}^{n - 1} \gets \text{SIMD\_SUB}\left(\langle \mathsf{s}_i \rangle_{i = 0}^{n - 1}, \langle \mathsf{t}_i \rangle_{i = 0}^{n - 1}\right)$ \\
 $\langle c_i \rangle_{i = 0}^{n - 1} \gets \text{SIMD\_SUB}\left(\langle \mathsf{s}_i \rangle_{i = 0}^{n - 1}, \langle \mathsf{p}_i \rangle_{i = 0}^{n - 1}\right)$ \\
 $\langle D_i \rangle_{i = 0}^{n - 1} \gets \text{SIMD\_ADD}\left(\langle D_i \rangle_{i = 0}^{n - 1}, \langle c_i \rangle_{i = 0}^{n - 1}\right)$ \\
 \Return $D=\sum^{n-1}_{i=0}2^{i \cdot \omega}D_i$ 

 \end{algorithm}

 Consider the addition of $A_i$ and $B_i$. The addition can be classified into one of the following three cases.  \hl{We refer to the case corresponding to the $i$-th lane as $L_i \in \{\bm{N},\bm{P},\bm{G}\}$.}
 \begin{itemize}
   \item Case $L_i=\bm{N}$: The addition would \textbf{N}ot generate a carry, which implies $c_{i+1} = 0$, regardless of $c_i$.
\item Case $L_i=\bm{P}$: The addition would \textbf{P}ropagate carry from previous limb, which implies $c_{i+1}= c_{i}$.
\item Case $L_i=\bm{G}$: The addition would \textbf{G}enerate a carry, which implies $c_{i+1} = 1$, regardless of $c_i$.
\end{itemize}
 Even without knowing the operands $A, B$, the carries for all limbs can be determined by the value of $L_i$.
 For example, when $(L_2, L_1, L_0) =(\bm{N},\bm{P},\bm{G})$, the carries $(c_3, c_2, c_1, c_0) = (0, 1, 1, 0)$ because 
 \begin{itemize}
     \item $L_0 = \bm{G}$ would give $c_1 = 1$,
     \item $L_1 = \bm{P}$ would propagate $c_1$ to $c_2$ and give $c_2 = c_1 = 1$,
     \item $L_2 = \bm{N}$ would give $c_3 = 0$.
 \end{itemize}

Algorithm~\textsf{ProposedAdd} employs this principle to transform a larger addition $A+B = (A_{n-1}, \dots, A_0) + (B_{n-1}, \dots, B_0)$ into a smaller one $\mathsf{t+p = (t_{n-1}, \dots, t_0) + (p_{n-1}, \dots, p_0)}$. \hl{Assume that the sum of $\mathsf{t}_i$ and $\mathsf{p}_i$ corresponds to the case $L_i'$. The values of $\mathsf{t}_i$ and $\mathsf{p}_i$ are determined such that $L_i' = L_i$.} This guarantees that the carry-overs from adding $\mathsf{t}_i$ and $\mathsf{p}_i$ align with those from adding $A_i$ and $B_i$.

Table~\ref{tab:cond} illustrates the process of converting a large addition to a small one. Each line in the ``Algorithm'' column of Table~\ref{tab:cond} corresponds to an assembly instruction. The first two rows of the table are similar, except for the carry-checking approach and the choice of $\mathsf{p}_i$. 

We opted for 8-bit integers for $\mathsf{p}_i$ and $\mathsf{t}_i$ because it is efficient to transfer data across 8-bit lanes in AVX-512 and SVE. While it is feasible to convert each limb to a 1-bit addition, we have not discovered an efficient method for such a conversion. Since the addition in Line 5 needs a cross-lane operation, we made $\mathsf{p}_i$ independent of $A_i$ and $B_i$. Consequently, $\mathsf{p}_i$ can be considered a constant and can be loaded from memory. 

We can show the correctness of the algorithm by the following lemmas and theorem.

\begin{lemma}
    The additions of $\mathsf{t}_i$ and $\mathsf{p}_i$ obtained from Table~\ref{tab:cond} falls into the same case as the additions of $A_i$ and $B_i$.
    \label{lem1}
\end{lemma}

\begin{proof}
\hl{We demonstrate that the additions of $\mathsf{t}_i$ and $\mathsf{p}_i$ align with the same case as $A_i$ and $B_i$, as shown in the third, fourth, fifth, and sixth columns of Table~{\ref{tab:cond}}. These columns represent the outcomes derived from each computation step for every instruction set and radix and for every case $L_i \in \{\mathbf{N}, \mathbf{P}, \mathbf{G}\}$.}
\end{proof}

\begin{lemma}
The carry $c_i$ obtained at Line 5 of Algorithm~\textsf{ProposedGenericRedc} is same as the carry $c_i$ in the addition of $A_i$ and $B_i$ \label{lem2}
\end{lemma}

\begin{proof}
\hl{By Lemma {\ref{lem1}}, the addition at Line 3 of the algorithm shares the same case as $A_i$ and $B_i$. The carry-over resulting from adding $\mathsf{t}_{i - 1}$ and $\mathsf{p}_{i - 1}$ to the sum of $\mathsf{t}_i$ and $\mathsf{p}_i$ matches the carry-over from adding $A_{i - 1}$ and $B_{i - 1}$ to the sum of $A_i$ and $B_i$. This carry-over is $c_i$. This implies that $\mathsf{s}_i = (\mathsf{t}_i + \mathsf{p}_i + c_i) \bmod 2^8$. The value of $c_i$ can be computed by taking $\mathsf{s}_i - \mathsf{t}_i - \mathsf{p}_i$, as indicated at Lines 4-5 of the algorithm.}
\end{proof}

\begin{theorem}
    Algorithm~\textsf{ProposedAdd} calculates the addition results of $A$ and $B$.
\end{theorem}

\begin{proof}
    By Lemma \ref{lem2}, the desirable value of $c_i$ is obtained at Line 5 of the algorithm. Then, $D_i = (D_i + c_i) \mod 2^{\omega} = (A_i + B_i + c_i) \mod 2^{\omega}$, which is the desirable result at the $i$-th limb at Line 6.
\end{proof}

The concepts of Algorithm~\textsf{ProposedAdd} are demonstrated in the subsequent example:
\begin{example}
Consider an example where $\omega = 16$, $n = 4$, $A_0 = 60000$, $B_0 = 5536$, $A_1 = 50000$, $B_1 = 15535$, $A_2 = 10000$, $B_2 = 10000$, $A_3 = 20000$, and $B_3 = 20000$. Assume that the numbers are represented using the native radix. Note that each limb here comprises 16 bits instead of the 64 bits assumed in Table \ref{tab:cond}. Hence, the values of 64, 65, and 191 from the table should be substituted with 16, 17, and 239 respectively for this example.

The sum of $A_0$ and $B_0$ invariably generates a carry. In contrast, the sum of $A_1$ and $B_1$ propagates a carry, while the sum of $A_2$ and $B_2$ does not generate a carry. Therefore, $(L_2, L_1, L_0)$ is $(\bm{N}, \bm{P}, \bm{G})$. Applying the SIMD\_ADD function from Line 1 of Algorithm~\textsf{ProposedAdd} yields $(D_3, D_2, D_1, D_0) = (40000, 20000, 65535, 0)$.

Using the native radix as a basis, the values of $\mathsf{t}_i$ and $\mathsf{p}_i$ are computed according to the first row in Table \ref{tab:cond}. This results in $(G_3, G_2, G_1, G_0) = (5, 5, 16, 0)$, $(m_3, m_2, m_1, m_0) = (0,0,0,1)$, and $(\mathsf{t}_3, \mathsf{t}_2, \mathsf{t}_1, \mathsf{t}_0) = (5, 5, 16, 17)$. The corresponding values for $(\mathsf{p}_3, \mathsf{p}_2, \mathsf{p}_1, \mathsf{p}_0)$ are then calculated to be $(239, 239, 239, 239)$.

Rewrite: The summation performed at Line 3 produces the outcome $(\mathsf{s}_3, \mathsf{s}_2, \mathsf{s}_1, \mathsf{s}_0) = (196, 197, 0, 0)$. After applying the SIMD\_SUB function at Lines 4-5, $(c_3, c_2, c_1, c_0) = (0, 1, 1, 0)$. This represents the expected carry for this addition. The final result derived from Line 6 is $(D_3, D_2, D_1, D_0) = (40000, 20001, 0, 0)$, which aligns with the expectations.
\end{example}

 \begin{table*}[h!]
\caption{Algorithms for operands related to 8-bit additions, which are capable of simulating 64-bit additions along with the outcomes}
\begin{center}
\begin{tabular}{cccccccc}
\hline
 \textbf{Radix}&\textbf{Architecture} & \multicolumn{4}{c}{\textbf{Operand 1 ($\mathsf{t}_i$)}}&\textbf{Operand 2} ($\mathsf{p}_i$)\\ 
~&~& Algorithm&$L_i=\bm{N}$ &$L_i=\bm{P}$ &$L_i=\bm{G}$&\hl{Algorithm}\\ 
 \hline
\makecell{Native\\Radix} &\makecell{AVX-512\\SVE}&\makecell{$G_i\gets \mathrm{popcnt}(D_i)$\\
								$m_i\gets (D_i<A_i)$\\
								$\mathsf{t}_i \gets \mathrm{mADD}(G_i,65,m_i)$}&
					\makecell{$0\le G_i<64$\\
								$m_i= 0$\\
								$\mathsf{t}_i<64$}&
					\makecell{$G_i=64$\\
								$m_i= 0$\\
								$\mathsf{t}_i=64$}&
					\makecell{$0\le G_i<64$\\
								$m_i= 1$\\
								$\mathsf{t}_i>64$}		
								&$\mathsf{p}_i \gets 191$\phantom{$-k$}\\ \hline
\makecell{Reduced\\Radix-$2^{k}$} &\makecell{AVX-512}&\makecell{$G_i\gets \mathrm{popcnt}(D_i)$\\
								$m_i\gets (D_i\ge 2^{k})$\\
								$\mathsf{t}_i \gets \mathrm{mADD}(G_i,65,m_i)$}&
					\makecell{$0\le G_i<k$\\
								$m_i= 0$\\
								$\mathsf{t}_i<k$}&
					\makecell{$G_i=k$\\
								$m_i= 0$\\
								$\mathsf{t}_i=k$}&
					\makecell{$0< G_i \le k$\\
								$m_i= 1$\\
								$\mathsf{t}_i>k$}		
								&$\mathsf{p}_i \gets 255-k$\\ \hline
\makecell{Reduced\\Radix-$2^k$} &SVE&\makecell{$G_i\gets \mathrm{SADD}(D_i, 2^{64}-2^k-1)$\\
								$\mathsf{t}_i\gets \mathrm{SSUB}(G_i, 2^{64}-3)$}&
								\makecell{$G_i<2^{64}-2$\\
								$\mathsf{t}_i=0$}&
								\makecell{$G_i=2^{64}-2$\\
								$\mathsf{t}_i=1$}&
								\makecell{$G_i=2^{64}-1$\\
								$\mathsf{t}_i=2$}&
								$\mathsf{p}_i \gets 254$\phantom{$-k$}\\ \hline
\multicolumn{8}{l}{Note: $\mathrm{popcnt}(D_i)=$Number of $1$ in binary representation of $D_i$. }\\
\multicolumn{8}{l}{\phantom{Note: }\hl{For this function, we utilize the VPOPCNTQ instruction in AVX-512 and the CNT instruction in SVE.}} \\
\multicolumn{8}{l}{\phantom{Note: }$\mathrm{mADD}(G_i, 65, m_i)=G_i+65$ $\bm{if}$ $m_i$ $\bm{else}$ $G_i$ }\\
\multicolumn{8}{l}{\phantom{Note: } \hl{For this function, we utilize the VPADDQ instruction in AVX-512 and the ADD instruction in SVE.}}\\
\multicolumn{8}{l}{\phantom{Note: }$\mathrm{SADD}(x, y)=\min(x+y, 2^{64}-1)$ \hl{For this function, we utilize the UQADD instruction in SVE.}}\\
\multicolumn{8}{l}{\phantom{Note: }$\mathrm{SSUB}(x, y)=\max(x-y, 0)$ \hl{For this function, we utilize the UQSUB instruction in SVE.}}\\
\end{tabular}
\label{tab:cond}
\end{center}
\end{table*}

\section{Proposed SIMD Montgomery Reduction}
\label{sec:red}
\subsection{Optimization for General Prime Numbers}

Algorithm~\textsf{ExistingGenericRedc} can be directly implemented with SIMD. However, Line 3 of the algorithm incurs significant computation cost due to its reliance on a cross-lane broadcast operation and a costly multiplication instruction, as documented in Table \ref{tab:lat}. 
Additionally, Line 4 in the algorithm is dependent on the completion of Line 3, and Line 3 must wait for the result of Line 4 in the previous iteration. These loop dependencies hinder the CPU pipeline's throughput, which is not a significant issue when using general instruction sets, where Line 4's computation time dominates. However, on SIMD, the computation time of Line 4 is much shorter, while Line 3's cost increases, making it a significant concern. 
To address this problem, we have  devised Algorithm~\textsf{ProposedGenericRedc}. 

\begin{algorithm}
\SetAlgorithmName{Algorithm~\textsf{ProposedGenericRedc}}{}{}
 \caption{Proposed SIMD generic Montgomery reduction}
 \label{alg:mont2}
 \KwIn {$T<pR$, $p  < 2^{\omega n}, r=2^\omega, R=2^{\omega n}, n>2$, $M_i \equiv r^{-n + i + 1} \pmod{p}$ for $1 \leq i \leq n - 2$, $p'\gets \frac{rr^{-1}-1}{p}$}
 \KwOut  {$TR^{-1} \bmod p$}
 Let $T_i$ be $\lfloor T/r^{i-1}\rfloor \bmod r$, i.e. $T = (T_{2n}, \dots, T_1)$\\
 $\langle H_i \rangle_{i = 1}^{n - 2} \gets \text{SIMD\_MUL\_nx1}\left(\langle M_i \rangle_{i = 1}^{n - 2}, \langle T_i \rangle_{i = 1}^{n - 2}\right)$ \\
 $T^{(n-2)}\gets \lfloor T/r^{n-2}\rfloor + \sum_{i = 1}^{n - 2} H_i$\\
  \For{$i\gets n-1$ \KwTo $n$}{
 $Q\gets T^{(i-1)}p'\bmod r$\\
 $T^{(i)}\gets (T^{(i-1)} +Qp)/r$
 }
   \uIf{$T^{(n)} >p$}{$T^{(n)} \gets T^{(n)}-p$}
      \uIf{$T^{(n)} >p$}{$T^{(n)} \gets T^{(n)}-p$}
 \Return $T^{(n)}$ 
 \end{algorithm}

 \hl{The function SIMD\_MUL\_nx1, found at Line 2 of Algorithm \textsf{ProposedGenericRedc}, computes the multiplication result of the $n$-limb integer $M_i$ and the one-limb integer $T_i$ for all $1 \leq i \leq n - 2$, deploying SIMD instructions. Considering the $n$ limbs of $M_i$ as $M_{i,0}, \dots, M_{i,n - 1}$, the multiplication outcome can be calculated via the following steps:} 
 \begin{enumerate}
     \item First, calculate 
     $$\langle {U}_j \rangle_{j = 0}^{n - 1} = \text{SIMD\_MUL\_H}\left(\langle M_{i,j} \rangle_{j = 0}^{n - 1}, \langle Z_j \rangle_{j = 0}^{n - 1}\right)$$ when $Z_j = T_i$ for all $0 \leq j \leq n - 1$;
     \item Similar to 1), calculate
     $$\langle {Y}_j \rangle_{j = 0}^{n - 1} = \text{SIMD\_MUL\_L}\left(\langle M_{i,j} \rangle_{j = 0}^{n - 1}, \langle Z_j \rangle_{j = 0}^{n - 1}\right)$$ when $Z_j = T_i$ for all $0 \leq j \leq n - 1$;
     \item Then, calculate $H_i = \sum_{j = 0}^{n - 1} U_j 2^{\omega (j + 1)} + \sum_{j = 0}^{n - 1} Y_j 2^{\omega j}$ using the algorithm \textsf{ProposedAddition} in Section \ref{sec:add}.
 \end{enumerate} 
Upon reaching Line 3 of Algorithm \textsf{ProposedGenericRedc}, the results of all $H_i$ are combined. Consequently, the addition in the third step can be postponed, leading to a further optimized computation within the algorithm.

To show the correctness of Algorithm~\textsf{ProposedGenericRedc}, let $T_i$ defined as in Line 1 of Algorithm~\textsf{ProposedGenericRedc}, $M_i \equiv r^{-n + i + 1} \pmod{p}$, $H_i = T_i M_i$, and $T^{(n-2)} = \lfloor T/ r^{n - 2}\rfloor + \sum_{i = 1}^{n - 2} H_i$. First, consider the following lemmas:
\begin{lemma}
$\textit{REDC}(T) \equiv T^{(n - 2)} r^{-2} \pmod{p}.$ \label{lem3}
\end{lemma}

\begin{proof}
We can derive that:
\begin{dmath}
    \textit{REDC}(T) \equiv TR^{-1} \equiv \left( \lfloor T/ r^{n - 2}\rfloor \cdot r^{n-2}+ \sum_{i = 1}^{n - 2} T_{i} r^{i - 1}\right)r^{-n}
    \equiv \left(  \lfloor T/ r^{n - 2}\rfloor  + \sum_{i = 1}^{n - 2} T_{i} r^{-n + i + 1} \right) r^{-2}
    \equiv \left(  \lfloor T/ r^{n - 2}\rfloor  + \sum_{i = 1}^{n - 2} T_{i} M_i \right) r^{-2}
    \equiv \left(  \lfloor T/ r^{n - 2}\rfloor  + \sum_{i = 1}^{n - 2} H_i\right)r^{-2}
    \equiv T^{(n - 2)} r^{-2} \pmod{p} \label{eqn2} 
\end{dmath}
\end{proof}
\begin{lemma}
    Let $T^{(n)}$ be the result obtained at Line 7 of Algorithm~\textsf{ProposedGenericRedc}. Then, $T^{(n)} \equiv REDC(T) \pmod{p}$.
    \label{lem4}
\end{lemma}

\begin{proof}
It is straightforward from the definitions that $T^{(n - 2)}$ can be obtained at Line 3. Then, Lines 4-7 are used for calculating $\textit{redc}\left(T^{(n - 2)},2\omega\right).$ By Lemma \ref{lem3},  
\begin{dmath} T^{(n)} \equiv \textit{redc}\left(T^{(n - 2)},2\omega\right) \hiderel{\equiv} T^{(n-2)}r^{-2} \hiderel{\equiv} REDC(T) \pmod{p}.
\end{dmath}
\end{proof}

\begin{lemma}
    Let $T^{(n)}$ be the result obtained at Line 7 of Algorithm~\textsf{ProposedGenericRedc}. Then, $T^{(n)} < 3p$. \label{lem5}
\end{lemma}
\begin{proof}
    From Line 3 of Algorithm~\textsf{ProposedGenericRedc}:
\begin{dmath}
T^{(n-2)} = \lfloor T/ r^{n - 2}\rfloor  + \sum_{i = 1}^{n - 2} H_i \hiderel{<} r^2p +(n-2)rp.\label{eqn1}
\end{dmath}
Using the similar argument as in (\ref{eqn:mont0}):
\begin{dmath}
    T^{(n - 1)} = \frac{T^{(n - 2)} + Qp}{r} < \frac{r^2p + (n-2)rp + (r - 1)p}{r} \hiderel{<} 2rp. \label{eqn:mont21}
\end{dmath}
Then,
\begin{dmath}
    T^{(n)} = \frac{T^{(n - 1)} + Qp}{r} \hiderel{<} \frac{2rp + (r - 1)p}{r} \hiderel{<} 3p. \label{eqn:mont22}
\end{dmath}
\end{proof}

\begin{theorem}
Let $T^{(n)}$ be the result obtained at Line 12 of Algorithm~\textsf{ProposedGenericRedc}. Then, $T^{(n)} = TR^{-1} \mod p$.
\end{theorem}
\begin{proof}
It is known from Lemma \ref{lem4} and Lines 8-11 that $T^{(n)} \equiv TR^{-1} \pmod p$. Also, because $T^{(n)}$ is less than $3p$ at Line 7, the value is less than $p$ by the final reduction at Lines 8-11. Hence, $T^{(n)} = TR^{-1} \mod p$. 
\end{proof}

When $T<pR-(n-2)r^{n-1}p$, it is worth noting that similar reasoning as in (\ref{eqn1})-(\ref{eqn:mont22}) can be employed to demonstrate that the value of $T^{(n)}$ obtained at Line 9 is less than $2p$. Therefore, the instruction at Lines 10-11 can be skipped. This condition holds for all the reductions in SIKE, CSIDH, and CTIDH, and hence, there is no need to execute the instruction at Lines 10-11 while implementing these cryptographic algorithms.

 Algorithm~\textsf{ProposedGenericRedc} requires additional memory reads, but this is not a problem as SIMD is efficient at memory reads and the load/write units in CPUs are not frequently used when computing Montgomery reduction. 
 
 Similar to Algorithm~\textsf{ExistingGenericRedc}, Algorithm~\textsf{ProposedGenericRedc} can be combined with large integer multiplication. Although $M_i$ is considered a pre-calculated constant, no information about $p$ is required to calculate it. Instead, it can be computed as $M_i=\textit{REDC}(r^{i-1})$ using Algorithm~\textsf{ExistingGenericRedc}. Therefore, the condition to use Algorithm~\textsf{ProposedGenericRedc} is the same as that for Algorithm~\textsf{ExistingGenericRedc}, and it can be applied to field arithmetic where the prime is variable, such as in RSA.
 
 The concepts of Algorithm~\textsf{ProposedGenericRedc} are illustrated in the subsequent example.
 \begin{example} \label{example2}
Set $\omega = 4$, $r = 2^4 = 16$, $n = 4$, $R = 2^{16} = 65536$, and $p = 62207$. Given that $(M_1, M_2) = (3888, 243)$, and $p'$ equals to $1$.

Suppose $T = 100000000$, which leads to $(T_8, \dots, T_1) = (0,5,15,5,14,1,0,0)$ as per Line 1 of Algorithm~\textsf{ProposedGenericRedc}. By Line 2, $H_1 = H_2 = 0$, and Line 3 gives $T^{(2)} = T/256 + 0 = 390625$.

In the loop's first iteration from Lines 4-7, $Q = 390625 \mod 16 = 1$ and $T^{(3)} = (390625 + 1 \times 62207)/16 = 28302$. By the second iteration, $Q = 28302 \mod 16 = 14$ and $T^{(4)} = (28302 + 14 \times 62207)/16 = 56200$.

As $T^{(4)} = 56200$ is already less than $p$, the final reductions at Lines 8-11 are not needed. Therefore, the result of this Montgomery reduction is $56200$.
 \end{example}

\subsection{Optimization for Primes in the Form $2^\ell F - 1$}
\label{sec:rsike}
We observe that the prime number $p$ used in several cryptographic systems such as SIKE \cite{NISTPQC-R4:SIKE22} or SQISign \cite{EPRINT:DeFLerWes22} can be written in the form of $p = 2^\ell F - 1$ when $F$ and $2^\ell$ are large integers. In this subsection, we aim to give an improvement over Algorithm~\textsf{ExistingSpecificRedc} for the case when $2^\ell \approx F$. In particular, assume that $\ell < \lceil n / 2 \rceil \omega <\ell + \omega$.

We are using the prime number $p_{503} = 2^{250}3^{159} - 1$ to demonstrate our idea, which can also be applied to other prime numbers. The prime number is often used in the SIKE cryptographic system \cite{NISTPQC-R4:SIKE22}. When $\omega = 64$, $p_{503}$ can be used as a 3-Montgomery-friendly modulus with $n=8$ limbs for $\omega=64$. In Algorithm~\textsf{ExistingSpecificRedc}, the PQCrypto-SIDH selected $m=2$ for $p_{503}$ because it is the largest divisor of $n$ that is not greater than $\lambda = 3$. This means that four multiplications between $Q$ (128 bits) and $M$ (256 bits) will be performed at lines 4-7 of Algorithm~\textsf{ExistingSpecificRedc}.

We aim to use the Karatsuba algorithm \cite{karatsuba1962multiplication} to speed up the multiplications. The Karatsuba algorithm allows us to reduce the complexity of multiplying two multiple-limb numbers, making it an attractive option for speeding up multiplications. \hl{It is discussed in \mbox{\cite{TCHES:CFGRR21}} that the algorithm can be efficiently implemented in SIMD.} However, the Karatsuba algorithm would be less effective if operands do not have the same number of limbs. Therefore, the Karatsuba algorithm cannot be used in the situation stated in the previous paragraph as the size of $Q$ and $F$ are not the same.

Our goal is to modify Algorithm~\textsf{ExistingSpecificRedc} by increasing the value of $Q$ to a larger number such as one with $\lceil n / 2 \rceil \omega$ bits. For example, when dealing with the prime number $p_{503}$, instead of using a $128$-bit value for $Q$ in Algorithm~\textsf{ExistingSpecificRedc}, it is more preferable to use a $256$-bit value for $Q$. \hl{It is not possible to achieve this by just simply replacing the value $2^{m\omega}$ in Lines 6-7 of Algorithm~\textsf{ExistingSpecificRedc} with a larger number. Merely replacing the component at the algorithm would result in an incorrect reduction output.} 

 \begin{algorithm}
 \SetAlgorithmName{Algorithm \textsf{ProposedSpecificRedc}}{}{}
 \caption{\hl{Proposed Montgomery reduction with $\lambda$-Montgomery-friendly modulus}}
 \label{alg:mont3}
 \KwIn {$T < pR, p=2^{\ell}F-1$ such that $\ell < \lceil n / 2 \rceil \omega <\ell + \omega, R=2^{\omega n}>p$}
 \KwOut  {$TR^{-1} \bmod p$}
 $T^{(0)} \leftarrow T$\\

  $t^{(1)} \gets [T^{(0)}F\bmod 2^{\omega}]\cdot 2^{\ell}$\\
 $Q^{(1)} \gets (T^{(0)}+t^{(1)})\bmod 2^{\lceil n/2\rceil\omega}$\\
 $T^{(1)}\gets \lfloor (T^{(0)}+2^{\ell} \cdot \text{Karatsuba}(Q^{(1)},F) ) /2^{\lceil n/2\rceil\omega} \rfloor$  \\ 
  \eIf{$n$ is odd number}{
  $Q^{(2)}\gets T^{(1)} \bmod 2^{\lfloor n/2\rfloor\omega}$\\
  $T^{(2)}\gets \lfloor (T^{(1)}+2^{\ell}\cdot \text{Karatsuba}(Q^{(2)},F))/2^{\lfloor n/2\rfloor\omega}\rfloor$
  }{
    $t^{(2)}\gets [T^{(1)}F\bmod 2^{\omega}]\cdot 2^{\ell}$\\
 $Q^{(2)}\gets (T^{(1)}+t)\bmod 2^{\lceil n/2\rceil\omega}$\\
  $T^{(2)}\gets \lfloor (T^{(1)}+2^{\ell}\cdot\text{Karatsuba}(Q^{(2)},F))/2^{\lceil n/2\rceil\omega} \rfloor$
  }
    \uIf{$T^{(2)} >p$}{$T^{(2)} \gets T^{(2)}-p$}
 \Return $T^{(2)}$ 
 \end{algorithm}

In order to yield a valid result for the reduction process, the variables $t^{(1)}$ and $t^{(2)}$ are introduced at the second and ninth line of Algorithm~\textsf{ProposedSpecificRedc}, respectively. Through the subsequent lemmas, we establish that by applying these corrections, Algorithm~\textsf{ProposedSpecificRedc} always provides a correct output. First, let us consider the variable $t^{(1)}$ at Line 2 and the variable $t^{(2)}$ at Line 9 of Algorithm~\textsf{ProposedSpecificRedc}. 

\begin{lemma}
    Let $r = 2^{\lceil n/2 \rceil \omega}.$ For $i \in \{1,2\}$, $t^{(i)} \equiv T^{(i - 1)}F2^\ell \pmod r$. \label{lem7}
\end{lemma}
\begin{proof} This lemma can be proved as follows:
\begin{dmath*} 
    t^{(i)} \bmod r = (T^{(i - 1)}F \mod 2^\omega) \cdot 2^\ell \mod 2^{\lceil n/2 \rceil \omega}
    = [(T^{(i - 1)}F \mod 2^\omega) \mod 2^{\lceil n/2 \rceil \omega - \ell}] \cdot 2^\ell
    = (T^{(i - 1)}F \mod 2^{\lceil n/2 \rceil \omega - \ell}) \cdot 2^\ell
    = T^{(i - 1)}F 2^\ell \mod 2^{\lceil n/2 \rceil \omega}.
\end{dmath*}
Thus, $t^{(i)} \equiv T^{(i - 1)}F2^\ell \pmod{r}$ 
\end{proof}

In the following lemma, let us consider the variable $Q^{(1)}$ at Line 3 and the variable $Q^{(2)}$ at Line 10 of Algorithm~\textsf{ProposedSpecificRedc}. 

\begin{lemma}
 Let $r = 2^{\lceil n/2 \rceil \omega}.$ For $i \in \{1,2\}$, consider $Q^{(i)}$ calculated at Line 3 and Line 10 of Algorithm~\textsf{ProposedSpecificRedc}. $T^{(i-1)} + pQ^{(i)}$ is divisible by $r$. \label{lem8}
\end{lemma}
\begin{proof} From Lemma \ref{lem7}:
\begin{dmath*}
    Q^{(i)} \equiv T^{(i - 1)} + t^{(i)} 
    \equiv T^{(i - 1)} + 2^\ell F T^{(i - 1)} 
    \equiv T^{(i - 1)} (2^\ell F + 1) \equiv T^{(i - 1)}(p + 2) \pmod{r}.
\end{dmath*}
Because $\lceil n/2 \rceil \omega < \ell + \omega \leq 2\ell$, we notice that $2^{2\ell}$ is divisible by $r = 2^{\lceil n/2 \rceil \omega}$. Hence, 
\begin{dmath*}
    T^{(i-1)} + pQ^{(i)} \equiv T^{(i - 1)} + p(p + 2) T^{(i - 1)}
    \equiv (p^2 + 2p + 1) T^{(i - 1)} 
    \equiv (p + 1)^2 T^{(i - 1)} \equiv 2^{2\ell}F^2 T^{(i - 1)} \equiv 0 \pmod{r}.
\end{dmath*}
\end{proof}

\begin{lemma}
The value $T^{(1)}$ calculated at Line 4 is $redc(T,\lceil n/2 \rceil\omega)$. The value $T^{(2)}$ calculated at Line 7 and Line 11 is $redc(T,n\omega)$. \label{lem9}
\end{lemma}
\begin{proof}
 Let $r = 2^{\lceil n/2 \rceil \omega}.$ From Lemma \ref{lem8}: 
\begin{dmath}
    \left\lfloor \frac{T^{(0)} + 2^\ell Q^{(1)} \cdot F}{r} \right\rfloor \equiv \left\lfloor \frac{T^{(0)} + Q^{(1)} (p + 1)}{r} \right\rfloor \equiv \left\lfloor \frac{T^{(0)} + Q^{(1)}p}{r} +  \frac{Q^{(1)}}{r}  \right\rfloor \equiv \frac{T^{(0)} + Q^{(1)}p}{r} + \left\lfloor \frac{Q^{(1)}}{r}  \right\rfloor 
    \equiv \frac{T^{(0)} + Q^{(1)}p}{r} 
    \equiv T^{(0)}r^{-1} \pmod{p} \label{eqn7}
\end{dmath}
Hence, $T^{(1)}$ is $redc(T^{(0)}, \lceil n/2 \rceil \omega)$. 

Next, let us consider the case when $n$ is odd. In that case, Lines 6-7 are used to calculate $T^{(2)}$. Let $r' = 2^{\lfloor n/2 \rfloor \omega}$. Then:
\begin{dmath}
    \left\lfloor \frac{T^{(1)} + 2^\ell Q^{(2)} \cdot F}{r'} \right\rfloor \equiv \left\lfloor \frac{T^{(1)} + Q^{(2)} (p + 1)}{r'} \right\rfloor \equiv \left\lfloor \frac{T^{(1)} + Q^{(2)}p}{r'} +  \frac{Q^{(2)}}{r'}  \right\rfloor \equiv \frac{T^{(1)} + Q^{(2)}p}{r'} + \left\lfloor \frac{Q^{(2)}}{r'}  \right\rfloor 
    \equiv \frac{T^{(1)} + Q^{(2)}p}{r'} 
    \equiv T^{(1)}(r')^{-1} \pmod{p}. \label{eqn8}
\end{dmath}
Hence, $T^{(2)}$ is $redc(T^{(1)}, \lfloor n/2 \rfloor \omega)$ and, since $\lfloor n/2 \rfloor \omega + \lceil n/2 \rceil \omega = n\omega$, $T^{(2)}$ is $redc(T^{(0)}, n\omega)$.

When $n$ is even, the same argument as in $(\ref{eqn7})$ can be used to show that $T^{(2)}$ is $redc(T^{(1)}, \lceil n/2 \rceil \omega)$  and, since $\lceil n/2 \rceil \omega + \lceil n/2 \rceil \omega = n\omega$, $T^{(2)}$ is $redc(T^{(0)}, n\omega)$.
\end{proof}

We are now ready to prove the correctness of Algorithm~\textsf{ProposedSpecificRedc} in the following theorem.
\begin{theorem}
The value $T^{(2)}$ returned at Line 15 of Algorithm~\textsf{ProposedSpecificRedc} is $TR^{-1} \bmod p$.
\end{theorem}
\begin{proof}
    By Lemma \ref{lem9}, it is left to show that $T^{(2)}$ at Line 15 is smaller than $p$. By the same argument as in (\ref{eqn7}):
    \begin{dmath}
        T^{(1)} \hiderel{=} \frac{T^{(0)} + pQ^{(1)}}{2^{\lceil n/2 \rceil \omega}} \hiderel{<} \frac{pR + p \cdot 2^{\lceil n/2 \rceil \omega}}{2^{\lceil n/2 \rceil \omega}} \hiderel{=} p2^{\lfloor n/2 \rfloor \omega} + p. 
    \end{dmath}
    Then, by the same argument as (\ref{eqn7}) and (\ref{eqn8}), at Line 13:
     \begin{dmath}
        T^{(2)} = \frac{T^{(1)} + p \cdot Q^{(2)}}{2^{\lfloor n/2 \rfloor \omega}} < \frac{p \cdot 2^{\lfloor n/2 \rfloor \omega} + p + p(2^{\lfloor n/2 \rfloor \omega} - 1)}{2^{\lfloor n/2 \rfloor \omega}} \hiderel{=} 2p. 
    \end{dmath} 
    By the final reduction at Lines 13-14, the result at Line 15 is smaller than $p$.
\end{proof}

Calculating the value of $t^{(i)}$ at Lines 2 and 9 is not resource-intensive. The computation only necessitates the least significant limb, which is derived from the product of the least significant limb of $T^{(0)}$ and that of $F$. Moreover, the least significant limb of $Q^{(i)} \cdot F$ at Lines 4 and 11 is also equal to $T^{(i - 1)} \cdot F \bmod{2^\omega}$. Consequently, the result of $t^{(i)}$ can be reused at those points. This means that the introduction of $t^{(i)}$ into Algorithm~\textsf{ProposedSpecificRedc} incurs no additional multiplication cost.

The bottleneck of Algorithm~\textsf{ProposedSpecificRedc} lies in the multiplications at Lines 4, 7, and 11. The Karatsuba method is employed to perform these multiplications. To demonstrate the efficiency of the proposed algorithm compared to Algorithm~\textsf{ExistingSpecificRedc}, we consider the multiplication of 512-bit integers. As previously discussed, the previous method requires four multiplications between 128-bit integers (2 limbs) and 256-bit integers (4 limbs), with each multiplication costing eight multiplication instructions. Therefore, a total of 32 multiplication instructions are needed. In contrast, the proposed method enables the reduction to be performed by two multiplications between 256-bit integers (4 limbs). If one-level Karatsuba method is used for these multiplications, each 256-bit (4 limbs) integer multiplication can be completed using 12 instructions, resulting in a total of 24 instructions. This reduces the number of SIMD multiplication instructions from 32 to 24, resulting in greater efficiency.

Primes of the form $2^\ell F - 1$ are commonly employed in isogeny-based cryptography. However, currently, only a few algorithms such as SIKE and SIDH use primes where $\ell < \lceil n / 2 \rceil \omega <\ell + \omega$. One other algorithm that uses such primes is Curve448, an elliptic curve cryptography algorithm which utilizes $p_{448}=2^{448}-2^{224}-1$. Unfortunately, a unique property of $p_{448}$ enables calculation of $\mathbb{F}_{p^{448}}$ without the need for Montgomery reduction. \hl{Despite the current circumstances, the employment of $p_{448}$ in elliptic curve cryptography signals the likelihood of more elliptic-curve based protocols using primes that can be represented as $2^\ell F - 1$ in the future.}

The concepts of Algorithm~\textsf{ProposedSpecificRedc} are illustrated in the subsequent example. The inputs of the following example is same as in Example \ref{example2}.

\begin{example}
  Set $\omega = 4$, $n = 4$, $R = 2^{16} = 65536$, $p = 62207 = 2^8 \cdot 243 - 1$, and $T = 100000000$. The value of $p$ allows us to deduce that $\ell = 8$ and $F = 243$.

At Line 3 of Algorithm~\textsf{ProposedSpecificRedc}, the calculation is made that $t^{(1)} = [(100000000 \times 243) \mod 16] \times 2^8 = 0$. Subsequent calculations yield $Q^{(1)} = 100000000 \pmod 256 = 0$ and $T^{(1)} = \lfloor 100000000 / 2^8 \rfloor = 390625$. Given that $n$ is an even number, Lines 9-11 of the algorithm come into play. This results in $t^{(2)} = [(390625 \times 243) \bmod 16] \times 2^8 = 768$. Consequently, the results obtained are $Q^{(2)} = (390625 + 768) \bmod 256 = 225$, and $T^{(2)} = \lfloor (390625 + 256 \times 243 \times 225)/256 \rfloor = 56200$, which accord with the output in Example \ref{example2}.
\end{example}

\section{Case Study and Experimental Results}
\label{sec:results}
This section will cover the implementation of prime field arithmetic for various cryptography algorithms and architectures. The benchmarks for ARM architecture are conducted on Wisteria/BDEC-01 (Odyssey) with A64FX@2.20GHz, while those for x64 architecture are run on Intel i7-1165G7 clocked at 2701 MHz. The benchmark for field arithmetic involves unrolling loops 80 times due to the calculation time being too short. The results are the average of 800,000 repetitions for field operations and 1,000 repetitions for the entire cryptography algorithm. We utilize the compiler's intrinsic to execute the calculation for AVX-512, whereas assembly code is employed for the tests on A64, x64, and SVE architectures.
\subsection {CTIDH with SVE}

\label{sec:rsve}
 \begin{table}[t]
\caption{Number of cycles of proposed SVE implementation for CTIDH-511 on A64FX}
\begin{center}
\begin{tabular}{cccc}
\hline
 \textbf{Operation}&\textbf{A64 (cycles)} & \textbf{SVE (cycles)}&\textbf{Speedup}\\ \hline
Addition&\phantom{00}16.07&\phantom{00}13.72&1.17x\\
Montgomery multiplication&406.98 &258.96 &1.57x\\
Scalar multiplication &\phantom{0}39.98 &\phantom{0}18.68 &2.14x\\
CTIDH Action &316,308,640 &242,948,411&1.30x\\  \hline
\end{tabular}
\label{tab2}
\end{center}
\end{table}
Our implementation of CTIDH utilizes SVE with native radix. While it is not ideal to assume the vector length of SVE, we have assumed a 512-bit implementation for optimal performance.

Although Algorithm~\textsf{ProposedAdd} has significantly improved the performance of addition on SVE for $\mathbb{F}_p$ addition and subtraction, it is 
 not significantly faster than A64 implementation. Therefore, we perform $\mathbb{F}_p$ addition and subtraction using A64. For $\mathbb{F}_p$ multiplication, we utilize Algorithms \textsf{ProposedAdd} and \textsf{ProposedGenericRedc}. The multiplication is carried out using an operand-scan technique, and the reduction step is implemented with Algorithm~\textsf{ProposedGenericRedc}. We use radix-$2^{56}$ for Lines 3-5 in Algorithm~\textsf{ProposedGenericRedc} instead of native radix, allowing us to perform the reduction with only one partial reduction in Lines 10-11, rather than two. This conversion of radix can be accomplished with a single table lookup instructions, since both radices is a multiple of 8. 

Since $p_{511}$ does not satisfy the condition for lazy correction, we still need to correct the result after reduction. However, a complete correction requires the carries for $T^{(n)}$ and $T^{(n)}-p$ to be both propagated, which we want to avoid since carry propagation is still costly even with Algorithm~\textsf{ProposedAdd}. We therefore only check for the most significant word of  $T^{(n)}$ to determine whether to use  $T^{(n)}-p$, and we postpone further correction until subtraction or equality testing becomes necessary. Given that subtraction and equality testing occur less frequently than multiplication, we believe that the tradeoff is worthwhile.

We evaluated the proposed method against the state-of-the-art CTIDH implementation proposed by Benegas et al.~\cite{TCHES:BBCCLMSS21}. However, since there is no existing SVE assembly code for their implementation, we adapted the assembly code for CSIDH by Jalali et al.~\cite{COSADE:JAKJ19} using the concept outlined in the CTIDH paper, and employed that code as our benchmark. \hl{We also implement the code of our algorithms based on that assembly code.}

As shown in Table \ref{tab2}, in SVE, the proposed \hl{SIMD} addition is 18\% faster than the standard A64 addition. We attribute this mainly to the benefit of SVE memory read-write capabilities. While A64 addition is a memory-bound function, the proposed SIMD addition is a compute-bound function. However, we observed a slight decrease in performance when using the proposed addition for CTIDH operations. This could be due to the fact that the remaining CTIDH operations are primarily compute-bound functions, and using A64 addition enables a better distribution of work between the compute execution unit and memory access execution unit.

Algorithm~\hl{\textsf{ProposedGenericRedc}} outperforms Jalali et al.'s Montgomery multiplication implementation by 57\%, and our scalar multiplication is faster than the current state-of-the-art implementation on SVE by 114\%. 
The speedup of our algorithm for CTIDH action is 30\%. Nonetheless, if our SIMD multiplication is utilized in conjunction with other algorithms that employ a prime of 510 bits or less, the resulting speedup would be even greater.

\subsection {CSIDH with AVX-512}
 \begin{table*}[t]
\caption{Performance results of proposed AVX-512 implementation for CSIDH-511 on TigerLake}
\begin{center}
\begin{tabular}{c|c|cc|cc}
\hline
  ~&\textbf{x64} &\multicolumn{2}{c|}{ \textbf{Previous AVX-512 Implementations}}&\multicolumn{2}{c}{ \textbf{Our AVX-512 Implementation}}\\ 
  \textbf{Operation}&\textbf{Number of cycles}&\textbf{Number of cycles}&\textbf{Speedup}&\textbf{Number of cycles}&\textbf{Speedup}\\ 
 \hline
\makecell{Native Radix Addition\\8 limbs}& \phantom{00}6.052 \cite{LC:CCCDRS19}&\phantom{0}11.593\phantom{*\cite{TCHES:CFGRR21}}&0.52x&\phantom{00}4.589\phantom{**}&1.32x\\ \hline
\makecell{Radix-$2^{43}$ Addition\\8 limbs}& \phantom{00}6.838\phantom{*\cite{LC:CCCDRS19}}&\phantom{00}7.672\phantom{*\cite{TCHES:CFGRR21}}&0.89x&\phantom{00}5.280\phantom{**}&1.30x\\ \hline
\makecell{Radix-$2^{43}$ Carry Propagation\\12 limbs, 2-packed}& \phantom{0}18.617*\phantom{\cite{LC:CCCDRS19}}&23.493*\cite{TCHES:CFGRR21}&0.79x&\phantom{0}10.627**&1.75x\\ \hline
\makecell{Montgomery Multiplication \\2-packed}&258.140 \cite{LC:CCCDRS19}&144.890 \cite{TCHES:CFGRR21} &1.78x&131.066\phantom{**}&1.97x\\ \hline
\makecell{Montgomery Squaring\\ 2-packed}&262.103 \cite{LC:CCCDRS19}&141.696 \cite{TCHES:CFGRR21} &1.85x&104.404\phantom{**}&2.51x\\ \hline
\makecell{CSIDH\\Action}&132,051,574 \cite{LC:CCCDRS19} &83,099,556 \cite{TCHES:CFGRR21}&1.59x&74,491,118&1.77x\\  \hline
\multicolumn{6}{l}{*\phantom{*}Input defined as (\ref{eqn:V})}\\
\multicolumn{6}{l}{**Input defined as (\ref{eqn:Vs}). Naive carry propagation using SSE.}
\end{tabular}
\label{tab9}
\end{center}
\end{table*}
This section will present the results obtained by utilizing AVX-512 with Algorithm~\textsf{ProposedAdd} and Algorithm~\textsf{ProposedGenericRedc}. We will compare the proposed methods' running time with two existing implementations, namely: 
\begin{enumerate}
    \item The state-of-the-art implementation of constant-time CSIDH on A64 by Cervantes-V{\'a}zquez et al. \cite{LC:CCCDRS19}, which employs the code from Castryck et al. \cite{AC:CLMPR18} for $\mathbb{F}_p$ arithmetic, and is based on the constant-time OAYT-style CSIDH implementation.
    \item The state-of-the-art implementation of constant-time CSIDH in SIMD by Cheng et al. \cite{TCHES:CFGRR21}, which is a constant-time OAYT-style CSIDH implementation on AVX-512.
\end{enumerate}
 It is worth mentioning that Cheng et al. \cite{TCHES:CFGRR21} have also optimized $\mathbb{F}_{p_{511}}$ squaring in their work, whereas Cervantes-V{\'a}zquez et al. utilized the code for multiplication to perform squaring.

Cheng et al. \cite{TCHES:CFGRR21} utilized a two-packed radix-$2^{43}$ representation to implement $\mathbb{F}_{p_{511}}$ arithmetic. This implies that an $\mathbb{F}_{p_{511}}$ element is represented with 12 limbs, and three vector registers are utilized to store the two operands. The storage order of each limb is defined as follows:
\begin{dmath} \bm{V} \hiderel{=}\langle a,b\rangle=\begin{Bmatrix}[a_0,a_3, a_6,a_{9\phantom{1}},b_0, b_3, b_6, b_{9\phantom{1}}]\\ [a_1,a_2, a_7,a_{10},b_1,b_4, b_7, b_{10}]\\ [a_2,a_5, a_8,a_{11},b_2,b_5, b_8, b_{11}]\end{Bmatrix}\label{eqn:V}
\end{dmath}


Our code for Algorithm~\textsf{ProposedGenericRedc} is based on Cheng et al.'s code \hl{\mbox{\cite{TCHES:CFGRR21}}}. In addition, we have performed some extra calculations to evaluate the performance of Algorithm~\textsf{ProposedAdd}. Table \ref{tab9} presents the benchmark results obtained from these calculations.

The top two rows of the table present the runtime performance of  Algorithm~\textsf{ProposedAdd} compared to the carry propagate addition implemented in x64 and AVX-512 when calculating one instance. The results demonstrate that the proposed algorithm's runtime on AVX-512 is significantly shorter than x64 and naive AVX-512 carry propagation. 

The third row shows that naive AVX-512 carry propagation is still slower than x64 for two-packed calculation. We encountered some difficulties while implementing Algorithm~\textsf{ProposedAdd} with the input defined in (\ref{eqn:V}). We found that the following storage order is better for SIMD implementation.
\begin{dmath} \bm{V^*}\hiderel{=}\langle a,b\rangle=\begin{Bmatrix}[a_0,b_0, a_3,b_3,a_6, b_6, a_{9\phantom{1}}, b_{9\phantom{1}}]\\ [a_1,b_1, a_4,b_4,a_7,b_7, a_{10}, b_{10}]\\ [a_2,b_2, a_5,b_5,a_8,b_8, a_{11}, b_{11}]\end{Bmatrix}
\label{eqn:Vs}
\end{dmath}
With this storage order, carries can be efficiently propagated using 128-bit SSE instructions without any expensive cross-lane operations. Row 3 has shown that it is very efficient to implement with input defined as (\ref{eqn:Vs}) using SSE. Although the experiment in Row 3 has an input loaded from cache, it is also efficient for SSE to calculate this carry propagation with input stored in AVX-512 registers. AVX-512 supports an instruction VEXTRACTI64x2, which allows CPU to read any 128-bit aligned lanes in an  AVX-512 registers to SSE registers. VEXTRACTI64x2 is a cross-lane instruction and have a CPI of 1 and latency of 3. Since these reads are independent to each other, SSE can handle input defined as (\ref{eqn:Vs}) stored in AVX-512 register efficiently.

Performance results for Montgomery multiplication are presented in the fourth row of the table, indicating an improvement of around 10\% over Cheng et al.'s implementation \cite{TCHES:CFGRR21}. While the proposed algorithm reduces the number of 52-bit multiplications by 6\%, we attribute the larger improvement in computation time because the proposed algorithm gives a more efficient CPU pipeline. The proposed algorithm can also enhance the performance of Montgomery squaring. Compared to the state-of-the-art algorithm, our implementation is 36\% faster in running time. This larger difference again confirms that the proposed Algorithm~\textsf{ProposedGenericRedc} have improved CPU pipeline. 

Row 6 of the table demonstrates that the proposed Montgomery multiplication and squaring contribute to an improvement in the CSIDH algorithm's performance. Our implementation is 11\% faster than Cheng et al.'s implementation and 77\% faster than Cervantes-V{\'a}zquez et al.'s x64 implementation.

\subsection{SIKE on ARM64}
In this section, we present our benchmark results for Algorithm~\textsf{ProposedSpecificRedc} using the \textit{SIKEp503} protocol parameter. This parameter set is widely recognized for its use in both SIKE and SIDH. \hl{This section takes into account the code from Microsoft's SIDHv3.5 \mbox{\cite{PQCryptoSIDH}}, a recognized highly optimized implementation of an isogeny-based cryptosystem. Modifications are performed to replace SIDHv3.5's addition and reduction operations with the newly proposed algorithms. The comparative results showcasing the latency difference before and after these alterations in SIDHv3.5 are presented in Table {\ref{tab1}}.}

 The proposed reduction method is 26\% faster than the reduction method of SIDHv3.5, which is close to the theoretical improvement where Comba has 33\% more multiplications than the one-level Karatsuba method. Regarding the overall performance of SIKE, our implementation is 7\% faster than SIDHv3.5. As we optimized only the reduction method and left many calculations of SIKE unchanged (e.g., multiplication), this result aligns with expectation.
 
\begin{table}[t]
\caption{Performance results of proposed Reduction for SIKE in comparison with SIDHv3.5 on A64FX}
\begin{center}
\begin{tabular}{cccc}
\hline
 \textbf{Operation}&\textbf{SIDHv3.5} & \textbf{new Reduction}&\textbf{Speedup}\\
&\textbf{Time (ns)}& \textbf{Time (ns)} \\ \hline
Reduction& 196.84&156.48&1.26x\\
Keygen&36,749,182 &35,204,915 &1.04x\\
Encapsulation &60,642,713 &56,449,034 &1.07x\\
Decapsulation &65,017,001 &60,901,455&1.07x\\  \hline
\end{tabular}
\label{tab1}
\end{center}
\end{table}

\subsection{Comparison with the Kogge-Stone Vector Addition}

We have not included a comparison between the Kogge-Stone vector addition \cite{avxadd} and the proposed addition algorithm in Tables \ref{tab2}-\ref{tab1}, primarily because the code provided in \cite{avxadd} pertains to additions with carry, whereas we focus on the addition without carry in these tables. Instead, we compare our proposed addition algorithm with previous works in this  section.

We modify the addition algorithm detailed in \mbox{\cite{avxadd}} for execution with SVE instructions. The average cycle count for the addition with carry operation stands at 55.28 cycles. However, our algorithm operates at a reduced count of 42.90 cycles. This performance difference reveals that our algorithm delivers a 29\% speed improvement in comparison to the  state-of-the-art addition algorithm in SVE. A reason which gives the proposed algorithm a better efficiency is the fact that the algorithm fixes the value of second operation ($\mathsf{p}_i$ in Table \ref{tab:inst}) to a constant. 

We have additionally carried out experiments employing AVX-512 instructions. The average cycle count observed for our algorithm stands at 6.41 cycles, in comparison to the Kogge-Stone vector addition which has an average of 4.16 cycles. The fact that our algorithm lags behind the state-of-the-art algorithm can be attributed to the relatively insignificant cost of transferring from mask registers to general-purpose ones in AVX-512. Therefore, our approach does not contribute a significant improvement to the execution time. 

\subsection{Discussions}

Our proposed addition and reduction algorithms have demonstrated their ability to enhance the performance of state-of-the-art SIMD algorithms in diverse scenarios. They have also shown the potential to boost the performance of post-quantum cryptographic protocols which rely on these operations. When compared to computations performed without SIMD (x64 or A64), the speed enhancements derived from our algorithm range from 1.17x to 2.51x. 

One could conjecture an improvement closer to 8x, given that SIMD allows for 8 simultaneous additions, subtractions, and multiplications. However, as illustrated in Table \ref{tab:inst}, the latency of SIMD instructions typically exceeds that of x64 or A64 instructions. The throughput (CPI) of AVX-512 is merely 2 - 4 times greater than that of x64. During our experimentation, we also observed that the improvement in SVE is even less noticeable, indicating that the expected speed up is unlikely to be less than 2x.  Additionally, SIMD requires the identical execution of eight parallel operations. Due to dependencies between operations, arranging them to fully harness the potential of SIMD is often impracticable. Considering these limitations, we believe that an improvement in the range of 1.17x to 2.51x is appreciable.

\section{Conclusions}
\label{sec:conclusions}
This research demonstrates that SVE can significantly enhance prime field arithmetic and cryptosystems such as CTIDH. However, we have identified that the existing algorithms for large addition and Montgomery reduction are not efficient for SVE on A64FX, resulting in too many pipeline stalls. To address this issue, we propose new algorithms for addition and Montgomery reduction, which lead to a 30\% speedup for CTIDH. These algorithms are not limited to SVE and have shown usefulness for AVX-512 as well. Additionally, we provide an algorithm for SIKE to improve its Montgomery reduction. Furthermore, we emphasize that addition should be implemented using general-purpose instructions or shorter SIMD to avoid the expensive cross-lane operation.

\bibliographystyle{IEEEtran}
\bibliography{abbrev3,crypto,biblio}
\begin{IEEEbiography}[{\includegraphics[width=1in,height=1.25in,clip,keepaspectratio]{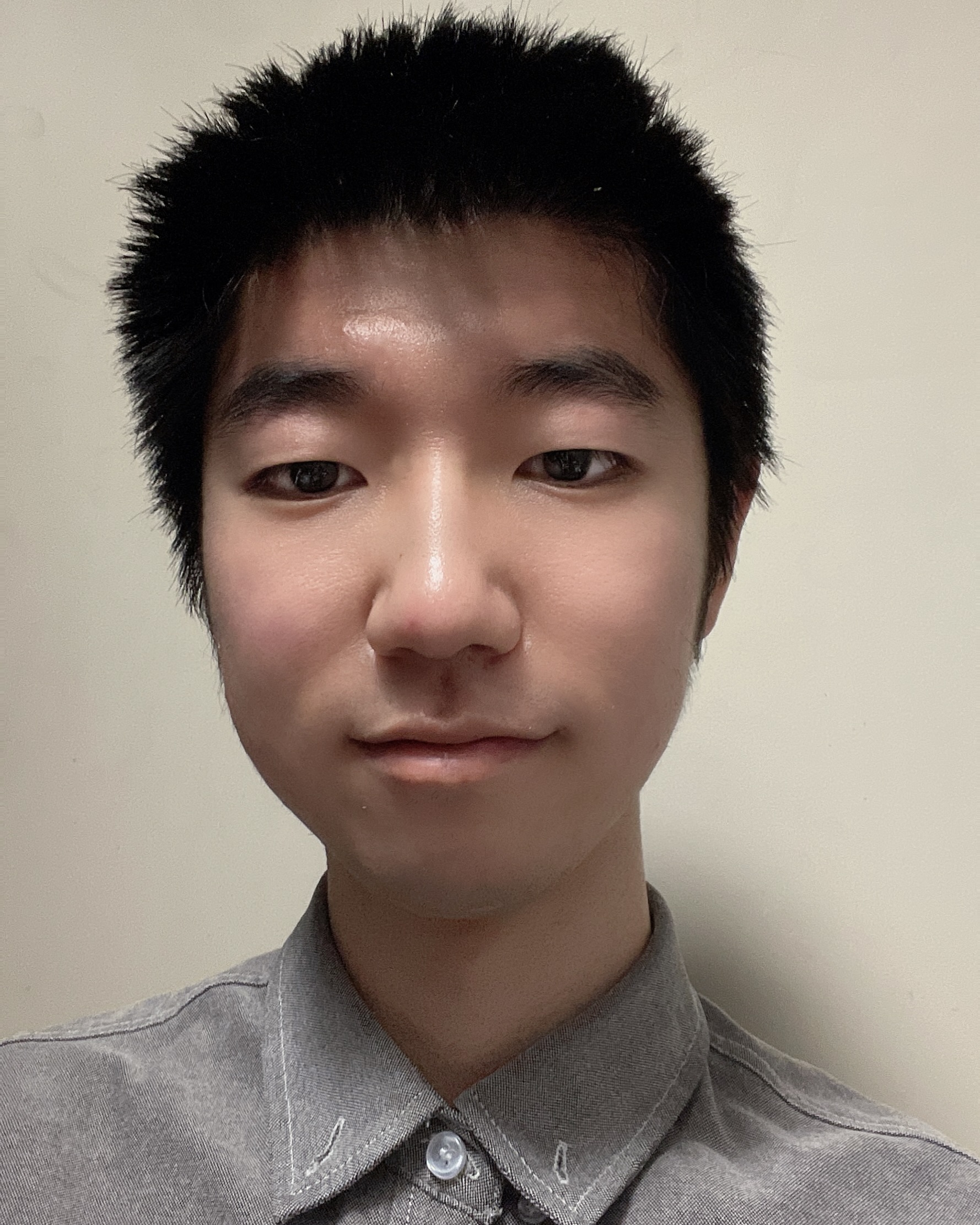}}]{Pengchang Ren} received the BE degrees in Chemical Engineering from East China University of Science and Technology and Osaka Prefecture University in 2022. He is currently pursuing his ME degree in Computer Science at the Graduate School of Information Science and Technology in The University of Tokyo. His research interests include parallel computing and program optimization.
\end{IEEEbiography}

\begin{IEEEbiography}[{\includegraphics[width=1in,height=1.25in,clip,keepaspectratio]{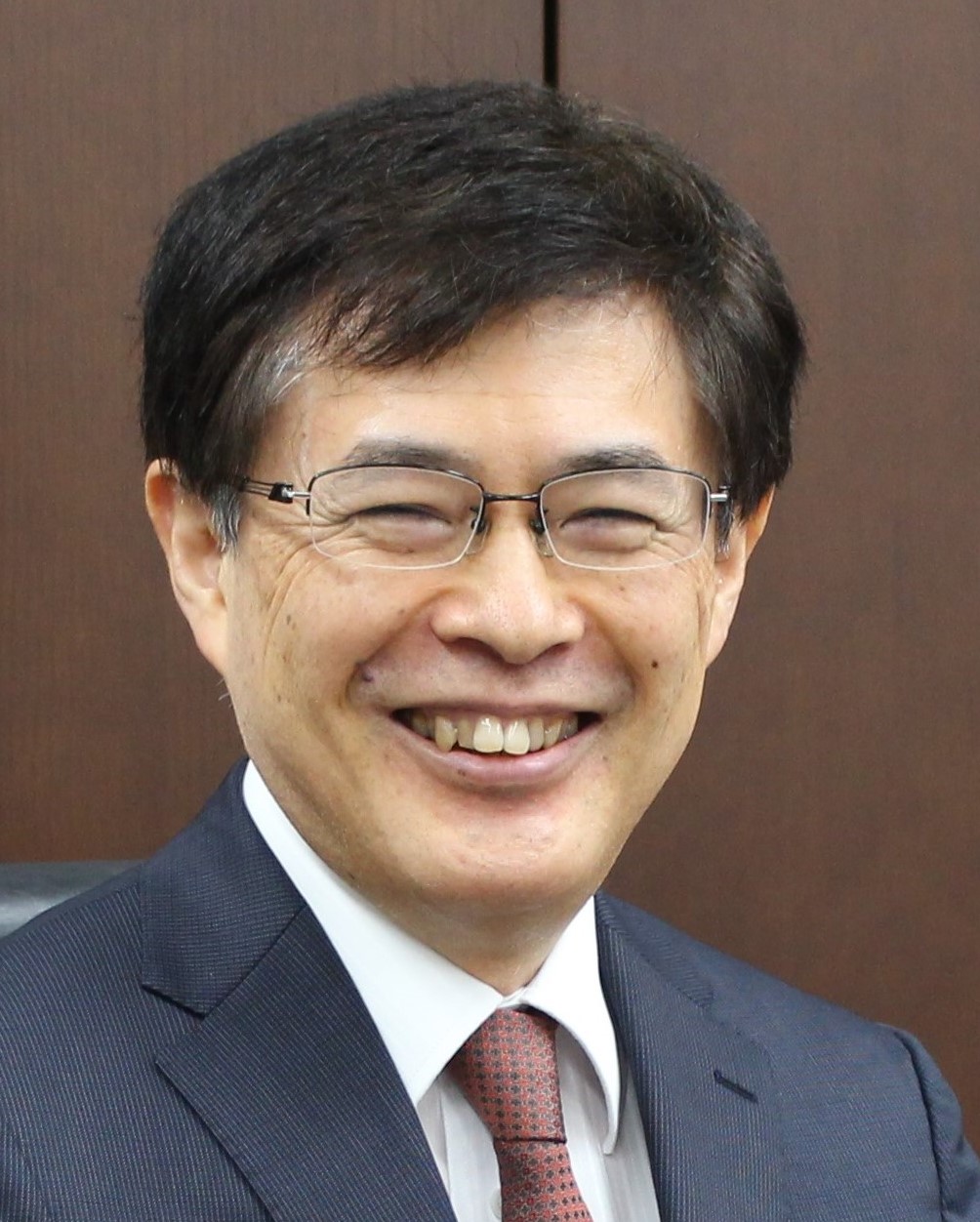}}]{Reiji Suda} graduated Master's course from Graduate School of Science, the University of Tokyo in 1993.  Then he worked as an research associate at the same department, after obtaining PhD degree, he moved to Nagoya University in 1997.  He returned to the University of Tokyo in 2002, and having been working as a professor since 2010.  His research topic is high performance computing and numerical algorithms.
\end{IEEEbiography}

\begin{IEEEbiography}[{\includegraphics[width=1in,height=1.25in,clip,keepaspectratio]{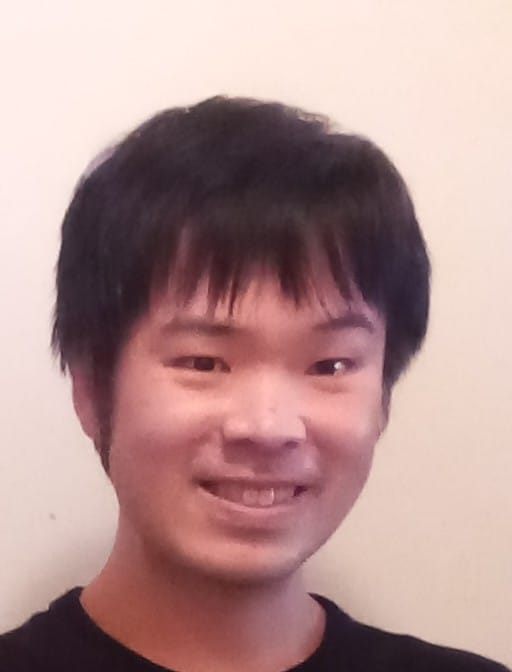}}]{Vorapong Suppakitpaisarn}
is currently an associate professor at the Department of Computer Science, Graduate School of Information Science and Technology at The University of Tokyo. He earned his Ph.D. from the same university in 2012. After graduation, he worked as a project lecturer at The University of Tokyo and the National Institute of Informatics before joining The University of Tokyo as an assistant professor in 2015. His research primarily focuses on applying combinatorial optimization techniques to reduce computation time in information security systems, particularly in the fields of elliptic curve and isogeny-based cryptography.
\end{IEEEbiography}

\end{document}